\documentclass[acmlarge]{acmart}
\AtBeginDocument{%
  }

\setcopyright{acmlicensed}
\copyrightyear{2018}
\acmYear{2018}
\acmDOI{XXXXXXX.XXXXXXX}

\usepackage{amsmath}
\usepackage{algorithmic}
\usepackage[linesnumbered,ruled,vlined]{algorithm2e}
\usepackage{soul}
\usepackage[normalem]{ulem}
\usepackage{tikz}
\usepackage{float}
\usepackage{enumerate}
\usepackage{multirow}
\usepackage{svg}
\usepackage{subfig}
\usepackage{hyperxmp}
\usepackage{color, colortbl}
\usepackage{paralist}

\definecolor{myblue}{HTML}{9ECAE1}

\newcommand{\cmmnt}[1]{}

\newcommand{\tinysection}[1]{\noindent \textbf{#1.}~}

\begin{document}

\title{Detecting Hard-Coded Credentials in Software Repositories via LLMs}


\author{Chidera Biringa}
\email{cbiringa@umassd.edu}
\orcid{0000-0001-5904-2764}
\affiliation{%
   \institution{University of Massachusetts Dartmouth}
   \streetaddress{285 Old Westport Rd, North Dartmouth, MA}
   \city{Dartmouth}
   \state{Massachusetts}
   \country{USA}
   \postcode{02747}
 }

 \author{G{\"o}khan Kul}
 \email{gkul@umassd.edu}
 \orcid{0000-0001-6467-1979}
 \affiliation{%
   \institution{University of Massachusetts Dartmouth}
   \streetaddress{285 Old Westport Rd, North Dartmouth, MA}
   \city{Dartmouth}
   \state{Massachusetts}
   \country{USA}
   \postcode{02747}
 }

\renewcommand{\shortauthors}{Biringa and Kul}


\begin{abstract}
Software developers frequently hard-code credentials such as passwords, generic secrets, private keys, and generic tokens in software repositories, even though it is strictly advised against due to the severe threat to the security of the software. These credentials create attack surfaces exploitable by a potential adversary to conduct malicious exploits such as backdoor attacks. Recent detection efforts utilize embedding models to vectorize textual credentials before passing them to classifiers for predictions. However, these models struggle to discriminate between credentials with contextual and complex sequences resulting in high false positive predictions. Context-dependent Pre-trained Language Models (PLMs) or Large Language Models (LLMs) such as Generative Pre-trained Transformers (GPT) tackled this drawback by leveraging the transformer neural architecture capacity for self-attention to capture contextual dependencies between words in input sequences. As a result,  GPT has achieved wide success in several natural language understanding endeavors. Hence, we assess LLMs to represent these observations and feed extracted embedding vectors to a deep learning classifier to detect hard-coded credentials. Our model outperforms the current state-of-the-art by 13\% $\in$ F1 measure on the benchmark dataset.
We have made all source code and data publicly available~\footnote{\texttt{https://github.com/PADLab/M2.git}} to facilitate the reproduction of all results presented in this paper.
\end{abstract}




\maketitle

\section{Introduction}
\label{sec:introduction}

The hard-coding of credentials such as passwords, cryptographic keys, and authentication tokens in distributed version control software development platforms such as GitHub creates attack surfaces exploitable by an adversary to conduct a backdoor attack~\cite{schuster2013towards},  privilege escalation~\cite{dagostino2021} or gain unauthorized access~\cite{bacon2008access}. Security researchers and practitioners have proposed several methods and approaches to detect these weaknesses, which fall under two independent but collaborative categories: (i) pattern-based~\cite{meli2019bad, sinha2015detecting, farinella2021git} and (ii) machine learning (ML)-based~\cite{credsweeper, lounici2021optimizing, feng2022automated, kall2021asynchronous} methods. In pattern-based methods, software repositories are investigated to identify high similarity patterns to a hard-coded credential~\cite{brodie2006scalable}. Although this approach has proven successful in well-defined cases~\cite{tatli2018password}, they detect established credential formats.   Hence, a minimal deviation in data format can generate edge cases necessitating human intervention~\cite{saleh2015method}. This is particularly common in today's world, with the rise in credentials heterogeneity and complexity, driven by organizations' desire to be more assertive in creating bottlenecks to reduce the occurrence of a data breach. The limitations posed by pattern-based approaches resulted in the application of ML to detect these weaknesses~\cite{feng2022automated}. 

LLMs powerful capacity to generalize to out-of-distribution observations, demonstrated in foundational models such as BERT~\cite{devlin2018bert} and GPT-2~\cite{radford2019language}, have led to successful adoptive approaches~\cite{hanif2022vulberta, rando2023passgpt} in tackling security problems. We leverage this strength in the representation of hard-coded credentials. While newer models like GPT-3~\cite{brown2020languagemodelsfewshotlearners}, Falcon~\cite{almazrouei2023falconseriesopenlanguage}, and LLAMA~\cite{touvron2023llama} offer advanced capabilities, we leverage BERT and GPT-2 due to their advantages in terms of computational efficiency, and model size and interpretability, making them more practical for integration into existing Continuous Integration (CI) pipelines. These models are well-documented, have extensive community support, and in our experiments, they provided close to perfect performance compared to previous system. For reference, in our assessments, the model size in a local deployment for BERT is approximately 1.34 GB and GPT-2 is 6 GB, whereas GPT-3 is 700 GB, Falcon 7B (smallest version) is 28 GB, and LLAMA-7B (smallest version) is 28 GB. For the purposes of fitting into a CI server, smaller model sizes would be considered better.

Given that credentials are generated using a concatenation of characters, we explore the capacity of LLMs pre-trained on large volumes of natural language data to represent these credentials. Consequently, we feed embeddings to a DL classifier to detect these vulnerabilities. Thus, this paper's goal is to detect hard-coded credentials via LLMs.

Concretely, the contributions of this paper are as follows: 
\begin{compactitem}

\item We introduce an approach for detecting hard-coded credentials by leveraging contextual embeddings derived from pre-trained language models. This method captures the nuanced syntactic and semantic features inherent in code, enabling precise classification of various credential types

\item Our experiments demonstrate our approach surpasses the performance of enterprise and research-oriented hard-coded credential detection tools, showing notable gains in overall detection accuracy.
\end{compactitem}

\subsection{Motivation}
Hard-coded credentials (CWE-798) are consistently ranked as one of the most dangerous software weaknesses by MITRE~\cite{mitre}. During software development, developers accidentally introduce credentials into the source code, configuration files, or documentation. This problem could lead to a significant data breach if the credentials go undetected. For example, in 2016, hackers stole the private data of 57 million Uber customers and drivers using login credentials found on a private area of GitHub~\cite{wong2017uber}. The adversaries used the stolen credentials to access Uber’s AWS account. The credential was accidentally introduced by one of its engineers. It reportedly cost Uber \$100,000 in ransom money paid to the hackers to delete the stolen data and conceal the breach. The handling of the leak resulted in the company firing the CSO. This occurrence highlights the severity and importance of this problem.

The problem becomes increasingly difficult and consequential with legacy software with millions of lines of code, increasing the complexity of detection. In this paper, we conduct experiments to answer the following research questions (RQ). Our approach leverages LLMs to learn the feature representations of hard-coded credentials in an unsupervised setting since the approach would be most useful in a CI environment to prevent developers from pushing the code that includes credentials into the repository. The learned input feature vectors for the credentials are extracted and propagated to a DL classifier for separability between hard-coded credentials, facilitating improved predictions.  

\begin{itemize}
\item [RQ1.] How to construct an LLM-assisted model for detecting hard-coded credentials? To answer this RQ, we experiment with masked and causal language modeling LLMs for feature representation and feed extracted embeddings to a standard DL classifier for detecting hard-coded credentials.

\item [RQ2.] What is the time cost of representing and detecting hard-coded credentials? To answer this RQ, we calculate the time it takes to represent and detect hard-coded credentials. 

\item [RQ3.] How does our model compare with other detection tools?
To answer this RQ, we evaluated our model against several other detection tools on a benchmark dataset.
\end{itemize}

\subsection{Scope and Taxonomy}
Hard-coded credentials (CWE-798) are the secret information accidentally introduced to software during development or maintenance. It ranks 18th spot in the 2023 top 25 most dangerous software weaknesses ~\cite{mitre}. Our model detects eight categories of these credentials: (i) passwords, (ii) generic secrets, (iii) private keys, (iv) predefined patterns, (v) seeds, salts, nonces, (vi) generic tokens, (vii) authentication keys and tokens, and (viii) others, as defined below. Note that the scope of this work does not include all other possible categories of hard-coded credentials such as development operations secrets and simple network management protocol community strings. Table~\ref{table:credexamples} shows examples of hard-coded credentials. Furthermore, undocumented and zero-day credentials are also not evaluated. Below are the definitions and the scope of our detection of hard-coded credentials.

\begin{table}[ht]
\centering
\caption{\texttt{Obfuscated Examples of Hard-coded Credentials}}
\vspace{-2mm}
\resizebox{0.99\linewidth}{!}{
\begin{tabular}
{{ p{0.20\linewidth} | p{0.85\linewidth} }}
\toprule
\texttt{Name} & \textbf{\texttt{Credential}} \\
\toprule
\texttt{Password} & \texttt{proxyServer.Certificate Manager . PfxPassword= "SsdGacriqvn"} \\
\midrule
\texttt{Generic Secret} & \texttt{SECRET\_KEY= '*fakqmj2o\_o3btm\%\ hbvoh1\$xfsd\_8nda(kf4x\-hl7k0gyh!5i4'} \\
 \midrule
 \texttt{Private Key} & \texttt{PEMBytes: []byte(`\-\-\-\-\-BEGIN RSA PRIVATE KEY\-\-\-\-\-} \\
\midrule
 \texttt{Predefined Pattern} & \texttt{accessKeyId: AKIAUNWKUPAVPRMGUUWX} \\
 \midrule
 \texttt{Seed Salt Nonce} & \texttt{oauth\_nonce: 'cBilmEAVPYhotlcTU59 eUCRb1tlDM1GGS7Fic6C3X'} \\
\midrule
 \texttt{Generic Tokens} & \texttt{token\_secret: 'cjdmmh3'} \\
 \midrule
 \texttt{Authentication Key and Token} & \texttt{oauth:\{consumer\_key: '5vvhj71r29jbo4c3'} \\
  \midrule
 \texttt{Other} & \texttt{user:w@/uzpmce@/} \\
\bottomrule
\end{tabular}
}
\label{table:credexamples}
\end{table}

\begin{itemize}
    \item Passwords are secret values that are generally considered the de-facto objects used in authenticating and granting access to a user of a digital account and assigning authorization privileges associated with that account~\cite{molloy2011attack}. They include a short secret with entropy less than 3.5 or a keyword stored in a variable name
    
    \item Generic Secrets are credentials that don't match a standard format like a password or API key. Instead, they show up as random, high-entropy strings—often generated on the fly using secure 
    random generators—which distinguishes them from more structured types~\cite{abdalla2009distributed}

    \item Private keys are the secret component of an asymmetric key pair. They enable one to decrypt data that’s been encrypted with the corresponding public key~\cite{delerablee2007identity}
    
    \item Predefined Patterns are credentials based on predefined established rules, which include Google API key --- distinctive identifiers used to verify a user or project~\cite{hu2013online} --- JSON Web Tokens --- standardized (RFC 7519) means of secure communication between parties using a JSON object~\cite{jones2015json} --- and Amazon Web Services client identifier. AWS client ID~\cite{cloud2011amazon} is a unique identifier that facilitates the authentication of a user's request to a server.
    
    \item Seeds, Salts, and Nonces are secrets with seed, salt, or nonce in a variable. Seeds are values with a 16-round Feistel structure of 128 bits input block size and key length~\cite{rogaway2004nonce}. A cryptographic salt is a randomly generated value primarily used to reinforce the security of one-way functions~\cite{rogaway2004nonce}. Nonce are one-off 4-bit hashed values~\cite{rogaway2004nonce}.

    \item Generic Tokens are digital objects used to authenticate the identity of an entity. Unlike more structured credentials like passwords or API keys, they’re often created on the fly to grant temporary access.~\cite{holmquist1999token}.
    
    \item Authentication Keys and Tokens are values to validate a user’s or entity’s identity~\cite{holmquist1999token}. They include credentials with "Auth" in the variable name and are not covered by preceding categories.

    \item Other refers to any hard-coded credential that doesn't clearly fit into one of our defined categories. It serves as a catch-all for ambiguous or irregular credentials that don't match specific formats like passwords or private keys
\end{itemize}

Lastly, we acknowledge that some researchers classify BERT and GPT-2 as PLMs but not LLMs (a subset of PLMs) due to lower number of parameters compared to newer LLMs, but this consideration does not bear any weight on the outcomes of this article as the techniques proposed still apply. 

\section{Language Architectures and Models}
\label{sec:background}
Our work leverages the ability of language models to capture unusual character combinations typical of hard-coded credentials, benefiting from the contextual understanding of transformer-based LLMs beyond semantics.

\textbf{Latent Semantic Analysis (LSA)}: An early semantic method using occurrence matrices and singular value decomposition to reduce dimensionality~\cite{deerwester1990indexing}, though limited by polysemy handling and positional context awareness.

\textbf{Feed Forward Language Models (FFLM)}: Improved upon LSA by using neural networks for distributed word embeddings~\cite{mikolov2013linguistic}, notably through models like continuous Bag-of-words~\cite{mikolov2013distributed} and GloVe~\cite{pennington2014glove}. However, they lack the capacity for sequential context understanding.

\textbf{Recurrent Neural Network Language Models (RNNLM)}: Designed for sequence modeling, RNNLMs consider previous outputs, addressing FFLM limitations~\cite{bengio2000neural}. Nevertheless, they face long-range dependency issues. Long Short-Term Memory Networks (LSTMs)~\cite{hochreiter1997long} and Gated Recurrent Units (GRUs)~\cite{bahdanau2014neural} address this by maintaining context over longer sequences, albeit with higher computational costs.

\textbf{Encoder-Decoder Language Models (EDLM)}: Use encoders for compression and decoders for sequence reconstruction~\cite{asadi2020encoder}. Transformer-based models significantly improved EDLM by employing self-attention, efficiently modeling long-range dependencies crucial for programming constructs and credential detection tasks~\cite{vaswani2017attention}.

Prominent transformer-based LLMs like Bidirectional Encoder Representations from Transformers (BERT)~\cite{devlin2018bert}, employing masked bidirectional language modeling, and Generative Pre-trained Transformers (GPT)~\cite{radford2019language}, using causal language modeling, have demonstrated powerful capabilities in capturing contextual nuances essential for effectively distinguishing hard-coded credentials.

\section{Approach}
\label{sec:methodology}
\begin{figure}[ht]
\centering
\includegraphics[width=0.90\linewidth]{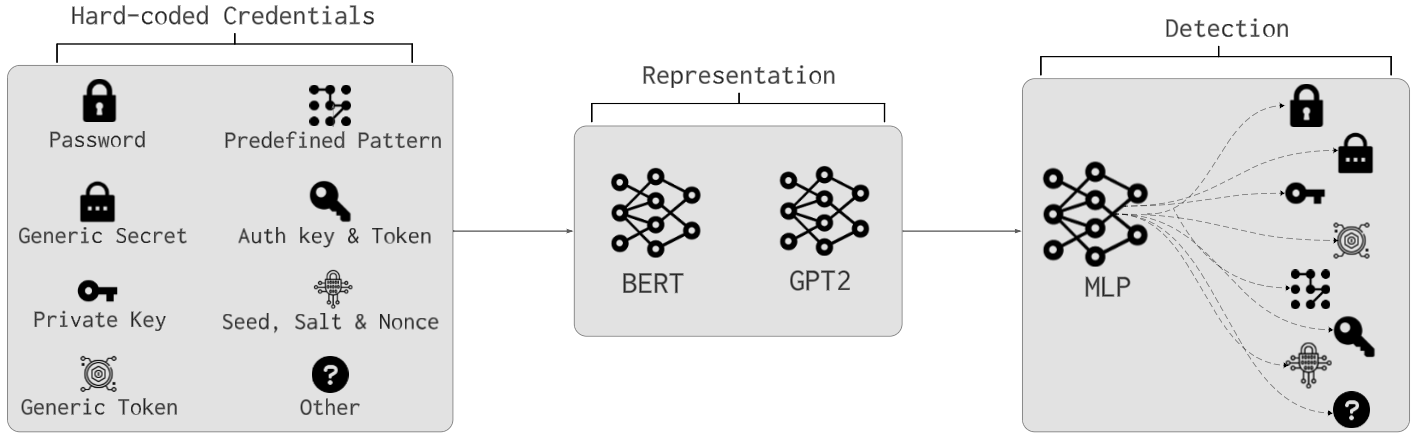}
\caption{Overview of Approach}
\label{fig:methodology}
\end{figure}

This section introduces our approach displayed in Figure~\ref{fig:methodology}. We detail the representation of hard-coded credentials using LLMs in Section~\ref{subsec:repr} and Section~\ref{subsec:dethcc} describes the implementation of our deep learning classifier for detecting hard-coded credentials. 

\subsection{Representing Hard-coded Credentials}
\label{subsec:repr}

Our representation process begins with the batching of input credential strings, which are then passed through a pre-trained transformer model. In this process, each batch undergoes tokenization using the model’s built-in tokenizer, which is tailored to the specific transformer architecture, whether it is BERT or GPT-2. The transformer then processes these tokens, generating token-level embeddings that capture the semantic information embedded within the input data.

We chose BERT and GPT-2 to represent hard-coded credentials because they often appear in natural language-like contexts, such as text files, config files, and comments, where natural language models are well-suited. While code-specific models such as CodeBERT~\cite{feng2020codebert} are trained on source code and have tokenizers tailored for programming languages, credential patterns frequently align naturally with general tokenization rather than strict code syntax. This makes general-purpose models effective in capturing credential structures across diverse formats.

After tokenization and embedding, we apply a combination strategy to aggregate the individual token embeddings into a single, unified vector representation for each credential. This strategy involves averaging the embeddings across the tokens, creating a fixed-length vector that encapsulates the entire credential. This vector acts as a compact, dense representation of the credential’s information.

Once the credential vectors are generated, we concatenate the embeddings from all batches into a single, comprehensive representation of the credentials. This unified representation is then fed into our downstream classifier for further processing. By leveraging pre-trained models like BERT or GPT-2, our approach effectively captures the contextual nuances inherent in hard-coded credentials, eliminating the need for extensive manual feature engineering and instead relying on the transformer’s ability to encode meaningful representations.

Hence, given the difficulty in distinguishing between different categories of hard-coded credentials, we leverage LLMs to learn real-valued vector representations of credentials in an unsupervised manner. By leveraging the generalization capabilities of LLMs, our approach effectively identifies hard-coded credentials that might be overlooked by traditional pattern-based methods. Additionally, the computational efficiency and interpretability of these models make them practical for integration into existing Continuous Integration (CI) pipelines, offering a scalable solution for credential detection.

\tinysection{LLMs Tokenization Strategies}
\label{subsection:toks}
Tokenization splits text into units to preserve syntax~\cite{biringa2022short}. We utilize WordPiece and Byte-level Encoding subword tokenization methods.

WordPiece, introduced by Kudo~\cite{kudo2018subword} and used by BERT~\cite{devlin2018bert}, preserves common words intact while decomposing infrequent words into smaller meaningful units. It starts from individual characters, applying merging rules to maximize symbol probability, thus reducing vocabulary size and improving data representation.

Byte-level Encoding, introduced by Sennrich~\textit{et al.}~\cite{sennrich2015neural} and utilized in GPT-2, restricts the vocabulary to 256 ASCII characters, pre-tokenizing text with standard delimiters. Frequent symbol pairs are merged iteratively, further reducing vocabulary size and enhancing representational efficiency.

\begin{figure}[ht] 
    \centering
    \captionsetup{justification=centering}
    \subfloat[BERT \newline Mean Intra-Class Distance: 39.958 \newline Mean Inter-Class Distance: 57.394 \newline Welch’s t-test: -827.806
    \newline p-value = 0.000e+00]
    {{\includegraphics[width=7cm]{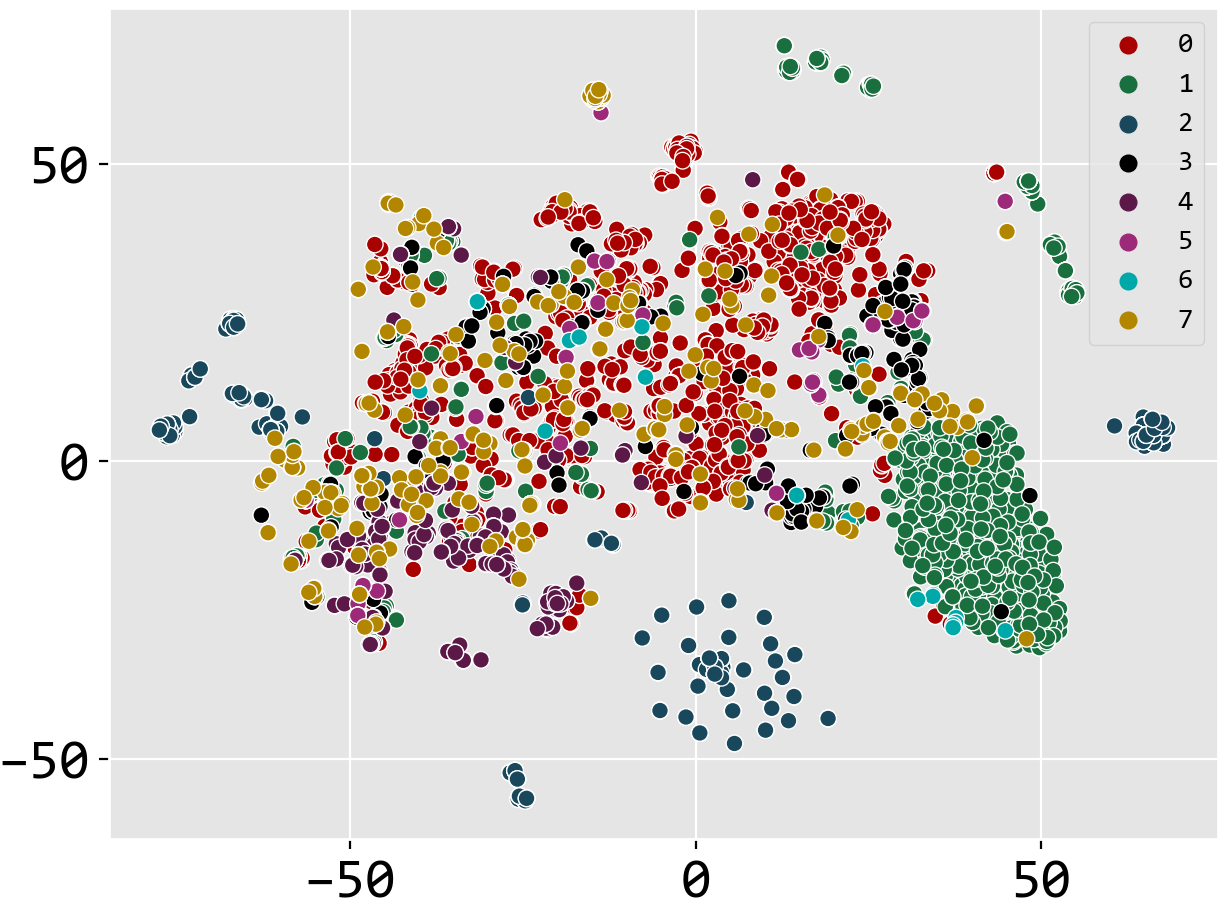}}}
    \;
    \subfloat[GPT2 \newline Mean Intra-Class Distance: 22.046 \newline Mean Inter-Class Distance: 60.432 \newline Welch’s t-test: t = -3226.507 \newline p-value = 0.000e+00] 
    {{\includegraphics[width=7cm]{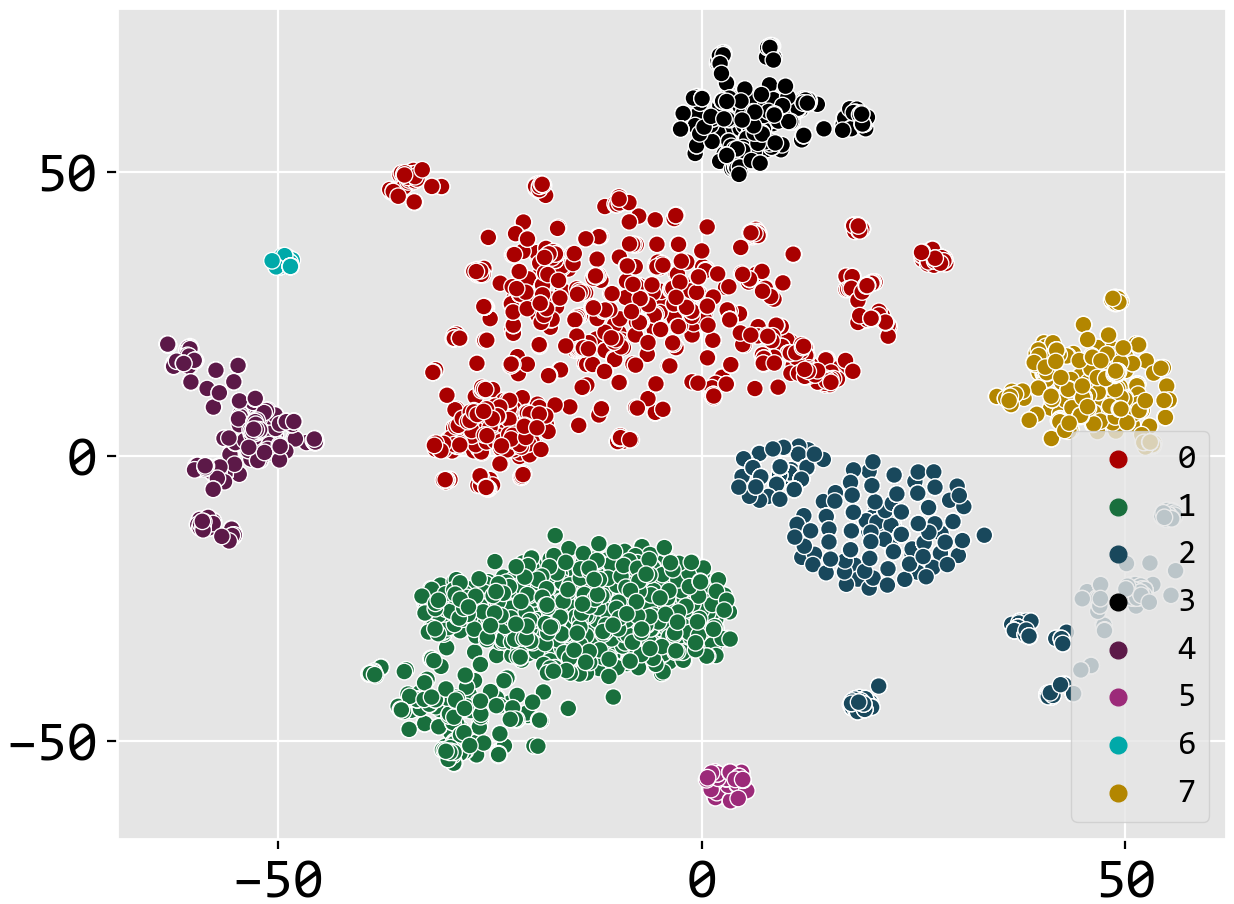}}}
    \caption{t-SNE Visualization of Represented Hard-Coded Credentials. 0: Passwords, 1: Generic Secrets, 2: Private Keys, 3: Generic Tokens, 4: Predefined Patterns, 5: Authentication Keys and Tokens, 6: Seeds, Salts, Nonces, 7: Others}
    \label{fig:tsne}
\end{figure}

We generate contextual embedding vectors for hard-coded credentials using pre-trained LLMs selected for their open accessibility and distinct learning approaches—BERT-base (mask language modeling, MLM) and GPT2 (causal language modeling, CLM). The dataset of hard-coded credentials is split into three subsets: 80\% for training, 10\% for validation, and 10\% for testing.

\textbf{Implementation Details:}
Using the Hugging Face Transformers and Simple Transformers libraries~\footnote{\url{https://shorturl.at/iDKQ4}}, we extract embeddings with the following hyperparameters: Learning Rate ($\alpha$): $4 \times 10^{-5}$, Adam Betas: $(0.9, 0.999)$, Adam Epsilon: $1 \times 10^{-8}$, Learning Rate Decay: $L2$ regularization, Optimizer: AdamW, Combination Strategy: Mean (for sub-token embeddings), and Batch Size: 32. We employ the mean pooling strategy (combine\_strategy = "mean") to obtain the final representation from both BERT and GPT2 models, averaging the hidden states across all tokens in a sequence. This ensures that the encoded representations capture the semantic content of the entire sequence, rather than relying on a single token, such as BERT's [CLS] or alternative sequence-wide pooling strategies. To mitigate overfitting, we train each model for 4 epochs—beyond which preliminary experiments showed performance degradation. After training, we extract the final hidden states, yielding 768-dimensional embedding vectors for each credential instance.

\textbf{Visualization and Analysis:}
We applied t-SNE~\cite{van2008visualizing} to visualize the high-dimensional embeddings. As illustrated in Figure~\ref{fig:tsne}, credentials belonging to the same category tend to form distinct clusters while dissimilar credentials are generally well-separated. Notably, embeddings from GPT2 appear to exhibit clearer inter and intra-category separation compared to those from BERT.

To substantiate these qualitative observations, we computed the pairwise Euclidean distances between credential embeddings and conducted Welch’s t-tests. We formulated the null hypothesis (H$_0$) that there is no difference between the mean Euclidean distances of embeddings from credentials of the same category (intra-class) and those from different categories (inter-class), against the alternative hypothesis (H$_1$) that the intra-class distance is significantly lower than the inter-class distance. For BERT, the mean intra-class distance was 39.958 and the mean inter-class distance was 57.394 (t = -827.8067, p $\approx$ 0), whereas for GPT2, the mean intra-class distance was substantially lower at 22.046 and the mean inter-class distance was higher at 60.432 (t = -3226.507, p $\approx$ 0). 

These low p-values indicate overwhelming evidence against H$_0$, confirming that GPT2’s embeddings are significantly more compact for the same credential type and more distinctly separated between different types. This robust statistical validation supports our claim that GPT2 effectively captures the latent features of hard-coded credentials, thereby enhancing downstream detection performance. These generated contextual embeddings are then used as input for our classification task to detect hard-coded credentials (see Section~\ref{subsec:dethcc}).

\subsection{Detecting Hard-coded Credentials}
\label{subsec:dethcc}
To detect hard-coded credentials, we implemented a multi-layer perceptron (MLP) classifier that operates on contextual embeddings generated by LLMs (i.e., \texttt{BERT-MLP} and \texttt{GPT2-MLP}). This feed-forward architecture was selected for its proven ability to separate latent representations effectively~\cite{rosenblatt1958perceptron}. The methodology is as follows:

\textbf{Data Preparation:}
The contextual embeddings were split into three subsets: 80\% for training, 10\% for validation, and 10\% for testing. This ensures that the model is trained on the majority of the data while preserving separate sets for hyperparameter tuning and final evaluation.

\textbf{MLP Architecture:}
Our MLP consists of: \begin{itemize} \item \textbf{Input Layer:} Accepts the embedding vectors. \item \textbf{Hidden Layers:} Three linear layers ($L_1$, $L_2$, $L_3$) are used. Each of the first two layers is followed by a rectified linear unit (ReLU) activation function (denoted as $A_1$ and $A_2$) and a dropout layer ($D_1$ and $D_2$) to mitigate overfitting. \item \textbf{Output Layer:} The final linear layer outputs raw scores for 8 credential categories, which are transformed into probabilities using a softmax function. \end{itemize} 

\textbf{Training and Evaluation Details:}
We trained the MLP using the Adam optimizer~\cite{kingma2014adam} with cross-entropy loss, a standard choice for classification tasks due to its effectiveness in handling probabilistic outputs. Hyperparameters were tuned using the validation set to ensure optimal performance. The \texttt{BERT-MLP} model was trained for 10 epochs, and the \texttt{GPT2-MLP} model for 4 epochs, both with a batch size of 32. These settings balance computational efficiency and model convergence. The learning rates were set to 1e-3 for \texttt{BERT-MLP} and 1e-4 for \texttt{GPT2-MLP}. After training, the model’s performance was evaluated on the test set to measure its generalization capability.

\section{Evaluation}
\label{sec:evaluation}
\tinysection{Hardware Setup} We conducted experiments on an Ubuntu 22.04.3 LTS, 128 CPUs, 1TB RAM, and an NVIDIA A100 80GB.

\begin{table}[ht]
    \centering
    \caption{CredData Dataset Statistics}
    \begin{tabular}{lrrr|lrrr}
        \toprule
        \textbf{Language} & \textbf{Total} & \textbf{Labeled} & \textbf{True} & \textbf{Language} & \textbf{Total} & \textbf{Labeled} & \textbf{True} \\
        \toprule
        Text & 85,144 & 10,896 & 1,626 & No Extension & 48,645 & 1,132 & 49 \\
        \midrule
        Go & 838,816 & 6,515 & 459 & Config & 7,920 & 340 & 46 \\
        \midrule
        YAML & 74,643 & 2,781 & 344 & AsciiDoc & 27,803 & 448 & 36 \\
        \midrule
        JavaScript & 742,704 & 4,130 & 340 & Shell & 42,019 & 1,340 & 31 \\
        \midrule
        Python & 351,494 & 5,643 & 260 & Haskell & 5,127 & 90 & 31 \\
        \midrule
        Markdown & 186,099 & 3,752 & 253 & Java Properties & 1,878 & 124 & 30 \\
        \midrule
        Java & 178,326 & 1,751 & 148 & reStructuredText & 38,267 & 531 & 21 \\
        \midrule
        Ruby & 186,196 & 4,669 & 145 & SQLPL & 16,808 & 612 & 20 \\
        \midrule
        Key & 8,803 & 364 & 116 & Objective-C & 19,840 & 211 & 14 \\
        \midrule
        TypeScript & 151,832 & 2,533 & 79 & TOML & 2,566 & 239 & 12 \\
        \midrule
        PHP & 113,865 & 1,936 & 76 & Scala & 9,564 & 163 & 12 \\
        \midrule
        JSON & 15,036,863 & 14,430 & 238 & Other & 1,226,683 & 7,803 & 172 \\
        \bottomrule
    \end{tabular}
    \label{table:language_stats}
\end{table}

\begin{table}[ht]
    \centering
    \caption{Credential Type Distribution in CredData Dataset}
    \begin{tabular}{lrr}
        \toprule
        \textbf{Credential Type} & \textbf{Count} & \textbf{Percentage} \\
        \toprule
        Passwords & 1,395 & 30.44 \\
        \midrule
        Generic Secrets & 1,056 & 23.04 \\
        \midrule
        Private Keys & 992 & 21.64 \\
        \midrule
        Generic Tokens & 333 & 7.27 \\
        \midrule
        Predefined Patterns & 327 & 7.13 \\
        \midrule
        Auth Keys and Tokens & 67 & 1.46 \\
        \midrule
        Salts, Seeds, and Nonces & 39 & 0.85 \\
        \midrule
        Others & 374 & 8.16 \\
        \bottomrule
    \end{tabular}
    \label{table:cred_types}
\end{table}

\tinysection{Dataset} 
To the best of our knowledge, CredData~\cite{sam_creddata} is the only publicly available hard-coded credential benchmark dataset~\footnote{\texttt{https://github.com/Samsung/CredData}}. CredData comprises eight hard-coded credential categories consisting of about 20 languages (Text, Go, YAML, JavaScript, Python, Markdown, JSON, Java, Ruby, TypeScript, PHP, AsciiDoc, XML, Shell, Haskell, reStructuredText, SQLPL, Objective-C, Scala and TOML), 19,459,282 lines of code, and 11,408 files.

The dataset was created by selecting publicly accessible GitHub repositories. It covers 181 topics across various languages, frameworks, and subjects. To build the dataset, they first gathered repositories from these topics and filtered for popularity using a minimum star count, resulting in 19,486 repositories. They then removed those with unsuitable licenses and identified candidates by searching for common credential-related keywords. Finally, they used open-source credential scanning tools and manually reviewed the results based on their labeling guidelines. 

Table~\ref{table:language_stats} presents the dataset's language statistics, highlighting the distribution of true credentials across programming languages. The \texttt{Text} category has the highest percentage at 35.5\%, followed by \texttt{Other} at 16.8\%. \texttt{Go} accounts for 10.0\%, while \texttt{YAML} and \texttt{JavaScript} contribute 7.5\% and 7.4\%, respectively. \texttt{Python} represents 5.7\%, \texttt{Markdown} 5.5\%, and \texttt{JSON} 5.2\%. \texttt{Java} and \texttt{Ruby} each account for 3.2\%.

Table~\ref{table:cred_types} presents the ground-truth labels of true credentials in the dataset totaled 4,583, which include passwords (30.44\%), generic secrets (23.04\%), private keys (19.64\%), generic tokens (21.65\%), predefined patterns (7.14\%), authentication keys and tokens (1.46\%), seeds, salts and nonces (0.85\%), and others (8.16\%). This distribution also highlights that our findings may be more robust for high-frequency categories such as passwords and generic secrets, compared to less common types like salts and nonces.

\tinysection{Performance Metrics} We evaluated the performance of predictive models using classical classification metrics: Accuracy, F1, Precision, Recall, and Matthews Correlation Coefficient (MCC). 


\begin{equation}
\begin{aligned}
\label{eqn:accuracy}
\displaystyle \texttt{Accuracy} = \frac{TP + TN}{TP + FP + FN + TN} 
\end{aligned}
\end{equation}

\begin{equation}
\begin{aligned}
\label{eqn:precision}
\displaystyle \texttt{Precision} = \frac{TP}{TP + FP} 
\end{aligned}
\end{equation}

\begin{equation}
\begin{aligned}
\label{eqn:recall}
\displaystyle \texttt{Recall} = \frac{TP}{TP + FN} 
\end{aligned}
\end{equation}

\begin{equation}
\begin{aligned}
\label{eqn:f1}
\displaystyle \texttt{F1} = \frac{2 \times (Precision \times Recall)}{Precision + Recall} 
\end{aligned}
\end{equation}

\begin{equation}
\begin{aligned}
\label{eqn:mcc}
\displaystyle {\texttt{MCC}}={\frac {{\mathit{TP}}\times {\mathit {TN}}-{\mathit {FP}}\times {\mathit {FN}}}{\sqrt {({\mathit {TP}}+{\mathit {FP}})({\mathit {TP}}+{\mathit {FN}})({\mathit {TN}}+{\mathit {FP}})({\mathit {TN}}+{\mathit {FN}})}}} 
\end{aligned}
\end{equation}

Recall denoted in Equation~\ref{eqn:recall} is the ratio of positive observations accurately predicted as positive observations, where FN denotes false negative. F1 denoted in Equation~\ref{eqn:f1}, computes the harmonic precision and recall mean performance, where $P$ is precision and $R$ is recall. MCC denoted in Equation~\ref{eqn:mcc}, measures the difference between actual and predicted observations. We introduced MCC to account for data imbalance. 

F1 score considers Precision and Recall together, but in the real-world, in the context of hard-coded secrets and credentials embedded in the code of the software, the trade-off between recall and precision is particularly important when considering the implications of false positives and false negatives. Recall, measures the ability of the detection system to identify all actual instances of hard-coded secrets. High recall is more significant because false negatives—instances where real secrets are not detected pose severe security risks. Undetected secrets can lead to data breaches, unauthorized access, and significant financial and reputational damage to organizations, as it has in the past. Therefore, maximizing recall to ensure that as many true positives as possible are identified is essential for robust security.

Precision, on the other hand, measures the accuracy of the detection system in identifying only true instances of hard-coded secrets. Low precision results in a high rate of false positives, where non-secret code is mistakenly flagged. The time and effort required to investigate false positives can slow down the development process and reduce overall productivity in software development.

In this context, recall is generally more impactful than precision because the primary goal is to ensure that no secrets go undetected, thus mitigating the risk of severe security breaches. However, maintaining a reasonable balance is essential to avoid excessive false positives, ensuring the detection system remains both effective and practical for developers to use.

\tinysection{Ethical Considerations} The creators of CredData obfuscated raw repository names, directory paths, and sensitive secrets. Furthermore, they notified the owners of public GitHub repositories used in creating the data. We have endeavored to maintain these ethics and only make publicly available credentials used in our experiments and not the raw repository data.

We perform our evaluation in the following subsections to answer the research questions defined.

\subsection{RQ1. How to construct an LLM-assisted model for detecting hard-coded credentials?}

In this RQ, we aim to ascertain the best LLM-assisted predictive model for detecting hard-coded credentials, thus we generate the input representations for hard-coded credentials using BERT and GPT2, and consequentially extract and feed embedding features as input vectors to an MLP classifier tasked with predicting hard-coded credentials. Table~\ref{table:rq1} (single run) displays the results of our experiments, \texttt{GPT2-MLP} is the best-performing model, outperforming \texttt{BERT-MLP} by 10\%, 26\%, 16\%, 28\% and 14\% on accuracy, F1, precision, recall, and MCC score. We argue that GPT2's ability to learn and segregate credentials in feature space as demonstrated in Figure~\ref{fig:tsne}, and facilitated by its significantly larger pretraining corpus is the primary contributing factor to its better performance.

\begin{table}[ht]
\centering
\caption{Performance Evaluation of Detecting Hard-coded Credentials using \texttt{BERT-MLP} and \texttt{GPT2-MLP}. for Single and Aggregated Runs}
\begin{tabular}{l|c|cc|c|cc}
\toprule
\multirow{3}{*}{\textbf{\texttt{Metric}}} & \multicolumn{3}{c|}{\textbf{\texttt{BERT-MLP}}} & \multicolumn{3}{c}{\textbf{\texttt{GPT2-MLP}}} \\  
\cline{2-7}
& \texttt{Single Run} & \multicolumn{2}{c|}{\texttt{Aggregated Runs}} & \texttt{Single Run} & \multicolumn{2}{c}{\texttt{Aggregated Runs}} \\  
\cline{2-7}
&  & \texttt{Mean} & \texttt{Std Dev} &  & \texttt{Mean} & \texttt{Std Dev} \\
\toprule
\texttt{Accuracy}  & 0.895  & 0.8728  & 0.006  & 0.997  & 0.996  & 0.002 \\
\midrule
\texttt{F1}        & 0.756  & 0.7640  & 0.011  & 0.985  & 0.973  & 0.012\\
\midrule
\texttt{Precision} & 0.850  & 0.8582  & 0.016  & 0.998  & 0.997  & 0.003 \\
\midrule
\texttt{Recall}    & 0.729  & 0.7386  & 0.011  & 0.975  & 0.959  & 0.016 \\
\midrule
\texttt{MCC}       & 0.866  & 0.8394  & 0.008  & 0.997  & 0.996  & 0.002 \\
\bottomrule
\end{tabular}
\label{table:rq1}
\end{table}

To further validate the robustness of our LLM-assisted predictive models for detecting hard-coded credentials, we conducted multiple training and evaluation runs using five different random splits of the dataset. For each model, we aggregated the performance metrics—accuracy, F1, precision, recall, and MCC—across these runs, reporting the mean and standard deviation. Table~\ref{table:rq1} presents the aggregated results for both \texttt{BERT-MLP} and \texttt{GPT2-MLP}. The results show that \texttt{GPT2-MLP} consistently outperforms \texttt{BERT-MLP}, achieving a higher mean accuracy (0.9969 vs. 0.8728), F1 score (0.9734 vs. 0.7640), precision (0.9972 vs. 0.8582), recall (0.9593 vs. 0.7386), and MCC (0.9962 vs. 0.8394). Additionally, the low standard deviations indicate minimal variance across different splits, demonstrating that \texttt{GPT2-MLP}’s superior performance is stable and not an artifact of a particular data partition. 

\subsection{RQ2. What is the time cost of representing and detecting hard-coded credentials?}

In this research question, we evaluate the latency involved in generating LLM representations of hard-coded credentials and the subsequent inference process. We measured the execution time for representing each credential category using both \texttt{BERT} and \texttt{GPT2} models, and computed the associated uncertainty estimates (i.e., standard deviation and 95\% confidence interval). The 95\% confidence interval, a standard measure of uncertainty, indicates a range within which the true mean is expected to fall~\cite{nakagawa2007effect}. Table~\ref{tab:timing_results} summarizes these results.

Our results indicate that, across all credential categories, \texttt{BERT} is substantially more efficient than \texttt{GPT2}. For example, in the case of \texttt{Passwords}, \texttt{BERT} achieves a mean execution time of 1.869 sec (Std Dev: 0.471 sec, 95\% CI: ±0.292 sec), compared to 2.788 sec (Std Dev: 0.071 sec, 95\% CI: ±0.044 sec) for \texttt{GPT2}. Similar trends are observed for other categories. Overall, \texttt{BERT} is approximately 1.871 times faster than \texttt{GPT2} in representing hard-coded credentials. This performance difference is consistent with the fact that the \texttt{GPT2} model contains roughly 1.39 billion more parameters than \texttt{BERT}, resulting in higher computational demands.

\begin{table}[ht]
\centering
\caption{Efficiency Evaluation of Representing Hard-coded Credentials with Uncertainty Estimation.}
\begin{tabular}{l|ccc|ccc}
\toprule
\multirow{2}{*}{\textbf{\texttt{Hard-coded Credential}}} & \multicolumn{3}{c|}{\textbf{\texttt{BERT (sec)}}} & \multicolumn{3}{c}{\textbf{\texttt{GPT2 (sec)}}} \\ 
\cline{2-7}
& \texttt{Mean} & \texttt{Std Dev} & 95\% \texttt{CI} & \texttt{Mean} & \texttt{Std Dev} & \texttt{95\% CI} \\ 
\toprule
\texttt{Passwords} & 1.869 & 0.471 & ±0.292 & 2.788 & 0.071 & ±0.044 \\
\midrule
\texttt{Generic Secrets} & 1.827 & 0.030 & ±0.018 & 3.265 & 0.103 & ±0.064 \\
\midrule
\texttt{Private Keys} & 0.950 & 0.035 & ±0.021 & 2.147 & 0.073 & ±0.045 \\
\midrule
\texttt{Generic Tokens} & 0.810 & 0.165 & ±0.102 & 2.082 & 0.168 & ±0.104 \\
\midrule
\texttt{Predefined Patterns} & 0.845 & 0.054 & ±0.033 & 2.319 & 0.158 & ±0.098 \\
\midrule
\texttt{Authentication Keys and Tokens} & 0.342 & 0.014 & ±0.009 & 1.736 & 0.107 & ±0.066 \\
\midrule
\texttt{Seeds, Salts, Nonces} & 0.317 & 0.041 & ±0.025 & 1.703 & 0.329 & ±0.204 \\
\midrule
\texttt{Others} & 0.793 & 0.055 & ±0.034 & 2.192 & 0.090 & ±0.056 \\
\bottomrule
\end{tabular}
\label{tab:timing_results}
\end{table}

\begin{table}[ht]
\caption{Efficiency Evaluation of Detecting Hard-coded Credentials with Uncertainty Estimation}
\centering 
\begin{tabular}{l|ccc}
\toprule
\textbf{\texttt{Model}} & \textbf{\texttt{Mean (sec)}} & \textbf{\texttt{Std Dev (sec)}} & \textbf{\texttt{95\% CI (sec)}} \\
\toprule
\texttt{BERT-MLP} & 0.0059 & 0.0002 & ±0.0001 \\
\midrule
\texttt{GPT2-MLP} & 0.0060 & 0.0002 & ±0.0001 \\
\bottomrule
\end{tabular}
\label{table:rq2b}
\end{table}

Given the sensitive nature of hard-coded credentials vulnerable to adversarial exploits, it is crucial to ensure that inference remains efficient with minimal performance overhead. Hence, we assess the efficiency of detecting hard-coded credentials in Table~\ref{table:rq2b}, incorporating uncertainty estimation to quantify variability in inference time.
We answer this research question by measuring the time taken to predict hard-coded credentials. \texttt{BERT-MLP} achieves a mean inference time of 0.0059 seconds (±0.0001 sec), while \texttt{GPT2-MLP} records a slightly higher mean inference time of 0.0060 seconds (±0.0001 sec). The minimal difference suggests that both models offer nearly identical inference efficiency, with negligible variance as indicated by their standard deviations and confidence intervals. It is important to note that the representation and inference times for hard-coded credentials are dependent on the hardware used in our experiments and should not be interpreted as universally applicable performance benchmarks.



\subsection{RQ3. How does our model compare with other detection tools?}
We answer this question by comparing our \texttt{GPT2-MLP} model to industry and enterprise-oriented detection tools. Given that \texttt{GPT2-MLP} is an ML-assisted approach, it isn't adequate to compare to other tools adopting traditional methods. However, we aim to present a robust evaluation of \texttt{GPT2-MLP} performance, so we include the results of other tools derived from CreData benchmark results~\cite{sam_creddata}. Hence, the primary focus of this research question is to compare \texttt{GPT2-MLP} against ML-assisted tools.

\begin{table}[ht]
\caption{Performance Evaluation of \texttt{GPT2-MLP} Against Other Hard-coded Credentials Detection Approaches. $^\dagger$ Aggregated mean F1 score over five random splits}
\centering
\begin{tabular}
{{ p{0.20\linewidth} | p{0.12\linewidth} | p{0.12\linewidth} | p{0.10\linewidth} | p{0.15\linewidth} }}
\toprule
\textbf{\texttt{Tool}} & \textbf{\texttt{Accuracy}} & \textbf{\texttt{Precision}}  & \textbf{\texttt{Recall}} & \textbf{\texttt{F1}} \\
\midrule
\texttt{Cred Sweeper} & 0.999 & 0.916 & 0.807 & 0.858 \\
\midrule
\texttt{Cred Digger} & 0.999 & 0.089  & 0.104 & 0.096 \\
\midrule
\texttt{Detect Secrets} & 0.999 &  0.141 & 0.381 & 0.206  \\ 
\midrule
\texttt{Git Leaks} &  0.999 &  0.525 & 0.244 & 0.333 \\
\midrule
\texttt{Shhgit} & 0.999 & 0.518 & 0.072 &  0.126 \\
\midrule
\texttt{Truffle Hog3} & 0.999 & 0.250 & 0.009 & 0.017 \\
\midrule
\texttt{Gitrob} & 0.999 & 0.224 & 0.195 & 0.209 \\
\midrule
\texttt{\textbf{GPT2-MLP}} & \texttt{\textbf{0.997}} & \texttt{\textbf{0.998}} & \texttt{\textbf{0.975}} & \texttt{\textbf{0.985}} \newline \texttt{\textbf{0.973 $\pm$ 0.012}}$^\dagger$ \\
\toprule
\end{tabular} 
\label{table:rq4} 
\end{table}

We first compare the performance of \texttt{GPT2-MLP} against non-ML-based tools, then against ML-based tools, and finally against both non-ML and ML-based tools. Table~\ref{table:rq4} presents our results, where \texttt{GPT2-MLP} outperforms non-ML-assisted tools, including \texttt{Detect Secrets}, \texttt{Git Leaks}, \texttt{Shhgit}, \texttt{Truffle Hog3}, and \texttt{Gitrob}, on the F1 measure by 130\%, 98\%, 154\%, 193\%, and 129\%, respectively. Tools such as PassFinder~\cite{feng2022automated} focus specifically on detecting passwords, whereas the other tools in our evaluation are designed for more comprehensive detection of various types of secrets. Outside of \texttt{GPT2-MLP}, \texttt{Git Leaks} performs the best, while \texttt{Truffle Hog3} performs the least effectively.

Additionally, \texttt{GPT2-MLP} surpasses ML-based tools, \texttt{Cred Sweeper} and \texttt{Cred Digger}, on the F1 measure by 13\% and 164\%, respectively. Notably, while the single-run F1 score for \texttt{GPT2-MLP} is 0.985, the aggregated mean F1 score over multiple random splits is 0.973 $\pm$ 0.012$^\dagger$, indicating consistently high performance with minimal variance. Finally, ranking the best-performing tools in order of effectiveness across both non-ML and ML-based approaches, we find \texttt{GPT2-MLP} at the top, followed by \texttt{Cred Sweeper}, \texttt{Git Leaks}, \texttt{Detect Secrets}, \texttt{Shhgit}, \texttt{Gitrob}, and \texttt{Truffle Hog3}.

\section{Limitations}
\label{sec:scope}
The method of using large language models (LLMs) such as BERT and GPT-2 for detecting hard-coded secrets in software shows promising potential for zero-day leak detection. Zero-day leaks refer to vulnerabilities that are exploited before they are known or addressed by developers. The key advantage of LLMs lies in their ability to generalize and identify patterns in data that were not explicitly present in their training set. This means that LLMs can potentially recognize new and unseen formats of secrets based on their learned understanding of language and code patterns. By leveraging the pre-trained embeddings and fine-tuning them on specific datasets related to secret detection, these models can identify anomalies and suspicious patterns indicative of hard-coded secrets, even if those patterns have not been previously documented or encountered.

However, it is important to acknowledge that this method may not detect all zero-day leaks. LLMs are limited by the data they were trained on, and their ability to generalize does not guarantee detection of entirely novel patterns that deviate significantly from what they have learned. Zero-day leaks often involve creative and unpredictable ways of embedding secrets, which can evade even sophisticated models. Moreover, the effectiveness of this method in zero-day leak detection depends on minimizing false negatives, as missing a new type of secret could lead to severe security breaches. While false positives are less critical in this context, a high rate can still hinder practical deployment due to developer fatigue. Overall, while LLMs offer a strong candidate for improving zero-day leak detection, relying solely on this method may not be sufficient. It should be complemented with other security practices and continuous refinement to enhance its detection capabilities.

Lastly, in our experiments, CredData has a modest data size despite being the only benchmark, to the best of our knowledge. Thus, the threat of a limited data size is the subject of future research. One avenue for approaching this limitation would be to expand the dataset by conducting a more robust scan beyond GitHub and other software management platforms to locate credentials that we can include in known categories and possibly identify new categories of credentials. Another approach would be to employ generative models such as generative adversarial neural networks to synthetically generate more data for experimentation. Furthermore, the primary limitation of our work is the lack of a methodical approach for detecting zero-day credentials hard-coded in software repositories in the wild.

\section{Related Work}
\label{sec:related_work}

\subsection{Pattern-Based}
\label{subsection:ptb}
Meli~\textit{et al.}\cite{meli2019bad} extensively investigated secret leakage in GitHub repositories. The authors analyzed over 100,000 repositories and billions of files scanned and monitored over six months. Their experiments showed that secrets are perpetually leaked on GitHub, highlighting the proliferation of vulnerability introduction in repositories. They also investigated possible root causes of the leak and proposed solutions to address the problem. 

Sinha~\textit{et al.}~\cite{sinha2015detecting} presented techniques potential adversaries adopt to identify and obtain accidentally introduced secrets, next the authors enumerated and evaluated a collection of techniques to detect and proactively prevent the introduction of API keys in version control management systems. 

Farinella~\textit{et al.}~\cite{farinella2021git} investigated the common pitfalls of detection mechanism in git servers as a service platform, followed by the proposal of an enterprise-oriented strategy that listens to organizations' public repositories for accidental introductions of secrets using detection tools such as TruffleHog. Positive credentials are acknowledged, and the responsible parties are notified. This approach facilitates rapid response, preventing the further entrenchment of the vulnerability to the code base.

Enterprise-oriented software security tools include TruffleHog~\footnote{\texttt{https://shorturl.at/23479}}, an open-source application security tool for detecting application programming interface keys and passwords. Yelp’s DetectSecret~\footnote{\texttt{https://shorturl.at/ksu37}} is an enterprise tool for secret detection. The tool is designed to be backward compatible with detecting and preventing the introduction of new secret variants and a means to migrate potentially highly malicious secrets to a more secure location.

Shhgit~\footnote{\texttt{https://shorturl.at/lnJO5}} is an open-source secret detection tool that continuously scans public GitHub, GitLab bitbucket, and local software repositories for usernames, API tokens, passwords, and private keys. Gitrob~\footnote{\texttt{https://shorturl.at/ahozL}} is an open-source tool to detect potentially sensitive files on public GitHub and GitLab repositories. It performs a deep scan of files in repositories, detects sensitive information, and presents results using a web interface to facilitate further analysis and investigation. GitLeaks~\footnote{\texttt{https://shorturl.at/dpU56}} is a SAST tool to detect and prevent the introduction of secrets to a repository.

\subsection{Machine Learning-Based}
\label{subsection:mlb}
CredSweeper~\footnote{\texttt{https://github.com/Samsung/CredSweeper}} is an enterprise-oriented tool for detecting hard-coded credentials in files. It starts by credential scanning in files and then applies matching rules to identify credential patterns before employing entropy, whitelist, and API validation filters used to detect the existence of credentials. Finally, it employs machine learning models to reinforce detection in the previous phase to reduce false positive predictions.

Feng~\textit{et al.}~\cite{feng2022automated} proposed PassFinder, an automated approach towards detecting passwords hard-coded in GitHub repositories. The authors conducted a longitudinal and large-scale experiment by monitoring password leakage of 539,012 repositories for 75 days. PassFinder utilizes deep neural networks to capture semantic and contextual features of textual passwords. The results of their experiments included the detection of 142,479 passwords from 64,045 of  539,012 investigated repositories, highlighting the severity and the relative ubiquitous nature of the threat.  

Lounici~\textit{et al.}~\cite{lounici2021optimizing} proposed an approach dubbed Credential Digger~\footnote{\texttt{https://github.com/SAP/credential-digger}} that leverages regular expressions and machine learning to detect hard-coded credentials from open source projects on GitHub. The tool scans repositories to identify hard-coded credentials and further deploys machine learning models to evaluate the aforementioned detected credentials with the aim of reducing the rate of false positive predictions.  The results of their experiments show an improved reduction in false positive detection. 

Kall and Trabelsi~\cite{kall2021asynchronous} proposed an asynchronous federated learning approach for hard-coded tokens and secrets. Federated learning is a privacy-preserving machine learning methodology that enables data training on local machines without data-sharing requirements. Thus, the authors tackled the problem of low hard-coded credential data availability due to privacy concerns from users and security professionals by locally trained data and then propagated to global state-of-the-art detection models for hard-coded credential detection.

Saha~\textit{et al.}~\cite{saha2020secrets} proposed a generalized framework combining regular expressions and machine learning to detect hard-coded secrets in code repositories, specifically addressing high false-positive rates. Using a Voting Classifier (Logistic Regression, Naïve Bayes, and SVM), their method achieved 84\% precision and 89\% recall across 300 GitHub repositories, effectively reducing false positives while maintaining robust secret detection.

Wen~\textit{et al.}~\cite{wen2022secrethunter} introduced SecretHunter, a scalable scanner using optimized metadata retrieval and reinforcement learning to efficiently detect secrets in public Git repositories. Their approach improved detection by 57\% over prior tools while significantly reducing bandwidth use. 

\section{Conclusion}
\label{sec:conclusion}
In this paper, we have presented a novel approach leveraging large language models (LLMs) such as BERT and GPT-2 for the detection of hard-coded credentials in software repositories. Our methodology capitalizes on the powerful feature representation capabilities of these models, enabling the identification of various types of hard-coded secrets with high accuracy. By combining these representations with a deep learning classifier, we demonstrated the effectiveness of our approach in detecting hard-coded credentials across multiple categories, including passwords, private keys, and tokens, exceeding the performance of the state-of-the-art significantly.

We showed that our approach achieves significant predictive accuracy, outperforming traditional pattern-based methods and providing competitive results compared to more recent detection tools. All the evaluation metrics we have used -- Accuracy, Precision, Recall, F1 Score, and Matthews Correlation Coefficient (MCC) -- highlight the robustness of our model in real-world scenarios. Particularly, our method's high recall underscores its capability to minimize false negatives, which is crucial for preventing security breaches caused by undetected secrets. The Recall for \texttt{GPT2-MLP} is 97.5\% compared to the closest tool Cred Sweeper's 80.7\%. We acknowledge that our approach may generate false positives despite the precision score of 99.8\%, as any other approach. This is higher than the Cred Sweeper's 91.6\%. Despite the lower difference here, the security implications of false negatives are far more severe. Therefore, our F1 score of 98.5\% compared to the closest approach's 85.8\% underscores the significant contribution that our approach has made.

Future work will focus on expanding our dataset to include a broader range of real-world software repositories and applications, thereby improving the generalizability and robustness of our model. Exploring the integration of our approach with existing development workflows and tools will also be a priority, ensuring seamless adoption by developers and security teams.

While newer models such as GPT-3, Falcon, and LLAMA offer advanced capabilities, we opted for BERT and GPT-2 due to their proven effectiveness, efficiency, and extensive community support. These models provide a balanced trade-off between performance and computational resources, making them practical for integration into CI environments. We would like to stress again that the parameter and model size of GPT-3, Falcon, LLAMA and similar LLMs would significantly increase the resource needs of a CI infrastructure, which commonly is not designed to run such workloads. However, based on the expanded dataset, we may consider testing the viability of such models, which would require significant resource investment from the CI infrastructure for this task, but the approach we built our work on will stay consistent with such models.

In conclusion, our LLM-assisted method for detecting hard-coded credentials represents a significant step forward in addressing this critical security challenge, offering a scalable and effective solution for protecting sensitive information in software development.

\begin{acks}
This research was supported in part by UMass Dartmouth's Marine and Undersea Technology (MUST) Research Program funded by the Office of Naval Research (ONR) under Grant No. N00014-23-1-2141. The views and conclusions expressed in this paper are those of the authors and do not reflect the official policy or position of the University of Massachusetts Dartmouth, the Office of Naval Research, U.S. Navy, U.S. Department of Defense, or U.S. Government.
\end{acks}

\bibliographystyle{ACM-Reference-Format}
\bibliography{reference}
\end{document}